\documentclass[a4paper, 11pt]{article}
\pdfoutput=1
\usepackage{jheppub}
\usepackage{afterpage}
\usepackage{multirow}
\usepackage{rotating}
\usepackage{epstopdf}
\usepackage{mathrsfs}
\usepackage{subfigure}
\usepackage{cancel}

\title{Axion dark matter with explicit Peccei-Quinn symmetry breaking in the axiverse}

\author{Hai-Jun Li}
\affiliation{Key Laboratory of Theoretical Physics, Institute of Theoretical Physics, Chinese Academy of Sciences, Beijing 100190, China}
\affiliation{Center for Advanced Quantum Studies, Department of Physics, Beijing Normal University, Beijing 100875, China}
\emailAdd{lihaijun@itp.ac.cn}

\abstract{
It was shown that the required high quality of the Peccei-Quinn (PQ) symmetry can be a natural outcome of the multiple QCD axions model.
In the axiverse, a hypothetical mass mixing between the QCD axions and axion-like particles (ALPs) can occur, which leads to an interesting phenomenon called the level crossing.
In this paper, we investigate this mass mixing between one QCD axion and one ALP with the explicit PQ symmetry breaking in the early Universe.
The dynamics of the axions and their cosmological evolutions when the level crossing occurs in this scenario are studied in detail.
We show the evolution of the mass eigenvalues and the mass mixing angle.
Then we check the condition for energy adiabatic transition with the corresponding parameter set.
Finally, we estimate the relic density of the QCD axion and ALP dark matter through the misalignment mechanism.
We find that, the QCD axion relic density can be suppressed, while the ALP relic density can be enhanced.
The level crossing in our scenario may have some cosmological implications, such as the axion domain walls formation, the nano-Hertz gravitational waves emission, and also the primordial black holes formation.}

\keywords{axions, cosmological phase transitions, dark matter theory, particle physics - cosmology connection}

\preprint{ITP-23-145}

\arxivnumber{2307.09245}

\begin{document}
\maketitle

\section{Introduction}

The nature of dark matter (DM) is a long-standing mystery in particle physics, cosmology, and astrophysics.
While the axion is one of the leading DM candidates, which was predicted by the Peccei-Quinn (PQ) mechanism with a spontaneously broken global $\rm U(1)_{PQ}$ symmetry \cite{Peccei:1977ur, Peccei:1977hh} to solve the strong CP problem in the Standard Model (SM), also called the QCD axion \cite{Weinberg:1977ma, Wilczek:1977pj, Kim:1979if, Shifman:1979if, Dine:1981rt, Zhitnitsky:1980tq}.
The QCD axion is a light pseudo-Nambu-Goldstone boson, which acquires a tiny mass from the QCD anomaly \cite{tHooft:1976rip, tHooft:1976snw}.
When the QCD instanton generates the QCD axion potential, the axion stabilizes at the CP conservation minimum value, which dynamically solves the strong CP problem.

As the DM candidate, the QCD axion can be non-thermally produced in the early Universe through the misalignment mechanism \cite{Preskill:1982cy, Abbott:1982af, Dine:1982ah}.
The QCD axion is massless at high cosmic temperatures, but as the temperature decreases, it gains a non-zero mass at the QCD phase transition and begins to oscillate when the mass is comparable to the Hubble parameter, which can explain the observed cold DM abundance.
The misalignment mechanism gives the upper limit of the classical axion window, $10^9\, {\rm GeV} \lesssim f_a\lesssim10^{12}\, {\rm GeV}$, where $f_a$ is the QCD axion decay constant, and the lower limit is given by the SN\,1987A neutrino burst duration \cite{Mayle:1987as, Raffelt:1987yt, Turner:1987by} and the cooling of the neutron star \cite{Leinson:2014ioa, Hamaguchi:2018oqw, Buschmann:2021juv}.
However, the origin of the PQ scale at this window is still unknown, which may come from the supersymmetry breaking scale and the Planck scale \cite{Carpenter:2009zs, Yin:2020dfn}.
In the case with $f_a\sim\mathcal{O}(10^{16})\, {\rm GeV}$, the axion DM abundance will be overproduced if without fine-tuning of the initial misalignment angle, which is the problem of overproduction of the QCD axion DM.

On the other hand, the PQ mechanism relies on the high quality of the global PQ symmetry, which works when the global PQ symmetry is explicitly broken and other explicit breaking effects are highly Planck-suppressed \cite{Kamionkowski:1992mf, Barr:1992qq, Ghigna:1992iv}.
There is a sharper argument that the continuous global symmetries should not exist in the quantum gravity theory \cite{Kallosh:1995hi, Banks:2010zn, Harlow:2018tng, Reece:2023czb}, and the PQ mechanism can be easily spoilt by the explicit PQ breaking operators \cite{Higaki:2016yqk, Jeong:2022kdr}.
From the low energy perspective, it is difficult to explain why the breaking of PQ symmetry other than QCD is so suppressed, which is the problem of high quality of the PQ symmetry.
The explicit breaking of PQ symmetry is strongly constrained by the neutron electric dipole moment (nEDM) measurement \cite{Abel:2020pzs}.
Since the QCD non-perturbative effects at high temperatures are suppressed, a small above breaking may have a significant impact on the axion dynamics.
Therefore, it is essential to investigate the explicit PQ symmetry breaking effects and their implications, and many works are studied. 
We are more concerned about a case that the required high quality of the PQ symmetry is a natural outcome of the multiple QCD axions model, $\rm e.g.$, the aligned QCD axion \cite{Choi:2014rja, Higaki:2014pja, Higaki:2014mwa}, and the clockwork QCD axion \cite{Choi:2015fiu, Kaplan:2015fuy, Long:2018nsl}.
In refs.~\cite{Higaki:2016yqk, Higaki:2015jag}, they show that the high quality PQ symmetry can be naturally explained in the aligned QCD axion models with the decay constants much smaller than the classical axion window. 
The implications of the explicit PQ symmetry breaking on QCD axion DM are also studied in refs.~\cite{Nakagawa:2020zjr, Jeong:2022kdr}.
Another effect of the explicit PQ symmetry breaking is reflected in the formation of QCD axion bubbles to generate the massive primordial black holes \cite{Kitajima:2020kig, Li:2023det, Li:2023zyc, Kasai:2023ofh}.
In addition, one possibility could be a suppression of the explicit PQ symmetry breaking due to their non-perturbative effects, see refs.~\cite{Alonso:2017avz, Hebecker:2018ofv} with the discussion of wormholes.

Several extensions of the SM, such as the string theory \cite{Green:1984sg, Witten:1984dg}, predicted the axion-like particle (ALP), which has similar properties to QCD axion but does not have to solve the strong CP problem.
In addition, the ALP mass and its coupling to the SM particles are not relevant, $\rm e.g.$, coupling to photon \cite{Halverson:2019cmy, Li:2020pcn, Cyncynates:2021xzw, Li:2022jgi, Li:2022pqa}, which are correlative in the QCD axion model.
The ALP is also the DM candidate and can be produced in the early Universe through the misalignment mechanism \cite{Cadamuro:2011fd, Arias:2012az, Chao:2022blc}.
The Universe with a large number of axions and ALPs is called the axiverse \cite{Svrcek:2006yi, Arvanitaki:2009fg, Demirtas:2021gsq}.
In the axiverse, the cosmological evolution called the level crossing can take place if there is a non-zero mass mixing between the QCD axions and ALPs, which are extensively studied in refs.~\cite{Hill:1988bu, Kitajima:2014xla, Daido:2015cba, Ho:2018qur, Daido:2015bva, Murai:2023xjn, Cyncynates:2023esj, Li:2023uvt}.
Generally, the condition for level crossing is that the zero-temperature mass and decay constant of the ALP are both much smaller than that of the QCD axion \cite{Ho:2018qur}.
This may induce the energy adiabatic transition of these axions, which is similar to the Mikheyev-Smirnov-Wolfenstein (MSW) effect \cite{Wolfenstein:1977ue, Mikheyev:1985zog, Mikheev:1986wj} in neutrino oscillations.
The adiabatic transition can lead to the suppression of the QCD axion energy density and isocurvature perturbations \cite{Kitajima:2014xla, Daido:2015cba}.
The ALP DM from the level crossing and the coupling to photon are studied in ref.~\cite{Ho:2018qur}.

In this paper, we investigate the non-zero mass mixing between the QCD axions and ALPs in the axiverse with the explicit PQ symmetry breaking in the early Universe.
Exactly, we consider the mixing between one QCD axion and one ALP.
The dynamics of the axions and their cosmological evolutions when the level crossing occurs in this scenario are studied in detail.
With the typical parameter set for level crossing to occur, we show the distribution of the mass eigenvalues and mass mixing angle, and also check the condition for energy adiabatic transition.
Next we estimate the relic density of the QCD axion and ALP DM through the misalignment mechanism.
We find that the QCD axion relic density can be suppressed, while the ALP relic density can be enhanced.
The level crossing in our scenario may also have some interesting cosmological implications.
 
The rest of this paper is organized as follows. 
In section~\ref{sec_Axion_dark_matter_review}, we first review the QCD axion and ALP DM, and introduce the mass mixing between them.
In section~\ref{sec_PQ_symmetry}, we investigate the explicit PQ symmetry breaking effects on axion mass mixing in the axiverse, focusing on the dynamics of the axions and their cosmological evolutions, and also the axion relic density. 
The conclusion is given in section~\ref{sec_Conclusion}.
 
\section{Axion mass mixing in the axiverse}
\label{sec_Axion_dark_matter_review}
 
In this section, we first review the QCD axion and ALP DM produced by the misalignment mechanism, and then briefly introduce the mass mixing between them.  

\begin{figure}[t]
\centering
\includegraphics[width=0.63\textwidth]{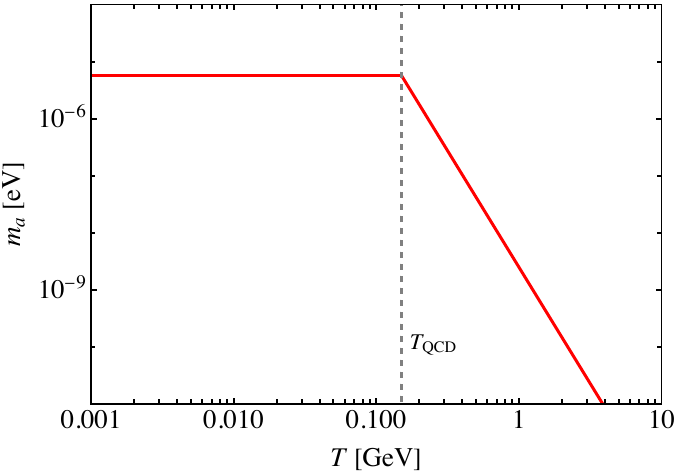}
\caption{The temperature-dependent QCD axion mass $m_a(T)$ as a function of the cosmic temperature $T$ for $f_a=10^{12}\,{\rm GeV}$.
The vertical gray line represents the temperature $T_{\rm QCD}$.}
\label{fig_maT}
\end{figure}
 
\subsection{QCD axion and ALP dark matter}

Here we consider the pre-inflationary scenario, $\rm i.e.$, the PQ symmetry is spontaneously broken during inflation.
The QCD axion ($a$) effective potential $V_{\rm QCD}(\phi)$ from the QCD non-perturbative effects is given by
\begin{eqnarray}
V_{\rm QCD}(\phi)=m_a^2(T) f_a^2\left[1-\cos\left(\dfrac{\phi}{f_a}\right)\right]\, ,
\label{sm_eq_Va}
\end{eqnarray}
where $\phi$ is the QCD axion field, $f_a$ is the QCD axion decay constant, $\theta=\phi/f_a$ is the QCD axion angle, and $m_a(T)$ is the temperature-dependent QCD axion mass 
\begin{eqnarray}
m_a(T)\simeq m_{a,0}\left(\dfrac{T}{T_{\rm QCD}}\right)^{-b}\, ,
\label{maT}
\end{eqnarray}
for the cosmic temperature $T\gtrsim T_{\rm QCD}\simeq150\, \rm MeV$ with $b\simeq4.08$ \cite{Borsanyi:2016ksw}.
Note that $m_{a,0}$ is the zero-temperature QCD axion mass \cite{GrillidiCortona:2015jxo}
\begin{eqnarray}
m_{a,0}=\dfrac{m_\pi f_\pi}{f_a}\dfrac{\sqrt{m_u/m_d}}{1+m_u/m_d}\simeq 5.70(7)\,{\mu \rm eV}\left(\dfrac{f_a}{10^{12}\,{\rm GeV}}\right)^{-1}\, ,
\label{ma0}
\end{eqnarray}
for $T< T_{\rm QCD}$, where $m_\pi$ and $f_\pi$ are the mass and decay constant of the pion, $m_u$ and $m_d$ are the up and down quark masses, respectively.
Using eqs.~(\ref{maT}) and (\ref{ma0}), we show the QCD axion mass as a function of the cosmic temperature in figure~\ref{fig_maT} for later use.
 
As the cosmic temperature decreases, the QCD axion starts to oscillate when its mass $m_a(T)$ becomes comparable to the Hubble parameter $H(T)$, which is given by the Friedmann equation in the radiation-dominated era
\begin{eqnarray}
3M_{\rm Pl}^2H^2(T)=\dfrac{\pi^2}{30}g_*(T)T^4 \, ,
\end{eqnarray}
where $M_{\rm Pl}\simeq 2.4\times10^{18}\, \rm GeV$ is the reduced Planck mass, and $g_*$ is the number of effective degrees of freedom of the energy density.
Using
\begin{eqnarray}
3H(T_{{\rm osc},a})=m_a(T_{{\rm osc},a})\, ,
\end{eqnarray}
then the oscillation temperature is given by
\begin{eqnarray}
T_{{\rm osc},a}\simeq0.96 \, {\rm GeV}\left(\dfrac{g_*(T_{{\rm osc},a})}{61.75}\right)^{-0.082}\left(\dfrac{f_a}{10^{12}\, \rm GeV}\right)^{-0.16}\, .
\label{Tosca}
\end{eqnarray}
The ratio of the QCD axion number density $n_a$ to the entropy density at $T_{{\rm osc},a}$ reads 
\begin{eqnarray}
\dfrac{n_a(T)}{s(T)}\bigg|_{T=T_{{\rm osc},a}}\simeq\dfrac{1}{2}\dfrac{m_a(T_{{\rm osc},a})}{s(T_{{\rm osc},a})}f_a^2\left\langle\theta_i^2f(\theta_i)\right\rangle\chi\, ,
\end{eqnarray}
where $s(T)=2\pi^2 g_{*s}(T)T^3/45$ is the entropy density, $g_{*s}$ is the number of effective degrees of freedom of the entropy density, $\theta_i$ is the QCD axion initial misalignment angle, $f(\theta_i)$ is the anharmonicity factor \cite{Lyth:1991ub, Visinelli:2009zm}
\begin{eqnarray}
f(\theta_i)\simeq\left[\ln\left(\dfrac{e}{1-\theta_i^2/\pi^2}\right)\right]^{7/6}\, ,
\end{eqnarray}
and $\chi\simeq1.44$ is a numerical factor that models the temperature-dependent QCD axion mass around $T_{{\rm osc},a}$ and depends on the number of quark flavors at $T_{{\rm osc},a}$ \cite{Turner:1985si}.
Then the current QCD axion DM abundance is given by
\begin{eqnarray}
\begin{aligned}
\Omega_ah^2&=m_{a,0} \dfrac{n_a(T)}{s(T)}\bigg|_{T=T_{{\rm osc},a}}\dfrac{s(T_0)}{\rho_{\rm crit} h^{-2}}\\
&\simeq0.14 \left(\dfrac{g_{*s}(T_0)}{3.94}\right)\left(\dfrac{g_*(T_{{\rm osc},a})}{61.75}\right)^{-0.42}\left(\dfrac{f_a}{10^{12}\, \rm GeV}\right)^{1.16}\left\langle\theta_i^2f(\theta_i)\right\rangle\, ,
\label{ogema_QCD_axion}
\end{aligned}
\end{eqnarray}
where $h\simeq0.68$ is the reduced Hubble constant, $T_0$ is the current cosmic microwave background (CMB) temperature, and $\rho_{\rm crit}=3H^2(T_0)M_{\rm Pl}^2$ is the critical energy density.
In order to explain the observed cold DM abundance, $\Omega_{\rm DM}h^2\simeq0.12$ \cite{Planck:2018vyg}, we have the initial misalignment angle
\begin{eqnarray}
\theta_i\simeq0.87\left(\dfrac{g_{*s}(T_0)}{3.94}\right)^{-1/2}\left(\dfrac{g_*(T_{{\rm osc},a})}{61.75}\right)^{0.21}\left(\dfrac{f_a}{10^{12}\, \rm GeV}\right)^{-0.58}\, .
\end{eqnarray}

The above misalignment mechanism can also be applied to the ALP DM.
The ALP can be directly extended from the QCD axion, which may be the low-energy consequence of the string theory \cite{Green:1984sg, Witten:1984dg}.
The string axiverse describes $N$ axion fields, which obtain the non-perturbative contributions to their collective potential from $M$ instantons with $M\gg N$ \cite{Arvanitaki:2009fg, Cyncynates:2021xzw}.
The effective potential is given by
\begin{eqnarray}
V_A(\psi_1,\dots,\psi_N)=\sum_{i=1}^M \Lambda_i^4\left[1-\cos\left(\sum_{j=1}^N \mathcal{Q}_{ij}\dfrac{\psi_j}{f_{A,j}}+\delta_i\right)\right]\, ,
\label{eq_VALP}
\end{eqnarray}
where $\psi$ is the ALP field, $f_A$ is the ALP decay constant, $\Lambda_i$ is the energy scale, $\mathcal{Q}_{ij}$ is the model parameter associated with the ALP charge, and $\delta_i$ is the constant phase.
In general, if considering the single ALP field $\psi$ with the constant axion mass $m_A=\rm const.$, the ALP DM abundance is given by 
\begin{eqnarray}
\begin{aligned}
\Omega_A h^2&=m_A \dfrac{n_A(T)}{s(T)}\bigg|_{T=T_{{\rm osc},A}}\dfrac{s(T_0)}{\rho_{\rm crit} h^{-2}}\\
&\simeq0.95 \left(\dfrac{g_{*s}(T_0)}{3.94}\right)\left(\dfrac{g_*(T_{{\rm osc},A})}{61.75}\right)^{-1} \left(\dfrac{m_A}{1\, \rm eV}\right)^{1/2}\left(\dfrac{f_A}{10^{12}\, \rm GeV}\right)^2\left\langle\theta_{i,A}^2f(\theta_{i,A})\right\rangle\, ,
\label{ogema_ALP}
\end{aligned}
\end{eqnarray}
where $n_A$ is the ALP number density, $T_{{\rm osc},A}$ is the ALP oscillation temperature that given by $3H(T_{{\rm osc},A})=m_A$, and $\theta_{i,A}$ is the initial misalignment angle of the ALP field.

\subsection{Axion mass mixing}

In this subsection, we briefly introduce the mass mixing between the QCD axion and ALP.
In the axiverse, it is natural to take into account the cosmological evolution of multiple axions.
In this case, the level crossing can occur if there is a hypothetical non-zero mass mixing between the QCD axions and ALPs \cite{Hill:1988bu, Kitajima:2014xla}.
Here we consider the mass mixing for the minimal scenario, one QCD axion $\phi$ and one ALP $\psi$, with the following low-energy effective Lagrangian
\begin{eqnarray}
\mathcal{L}\supset\dfrac{1}{2}\partial_\mu\phi\partial^\mu\phi +\dfrac{1}{2}\partial_\mu\psi\partial^\mu\psi-V_{\rm mix}(\phi,\psi)\, ,
\end{eqnarray}
where $V_{\rm mix}(\phi,\psi)$ is the mixing potential \cite{Kitajima:2014xla}
\begin{eqnarray}
V_{\rm mix}(\phi,\psi)=m_a^2(T) f_a^2\left[1-\cos\left(\dfrac{n_1\phi}{f_a}+\dfrac{n_2\psi}{f_A}\right)\right]+ m_A^2 f_A^2\left[1-\cos\left(\dfrac{n_3\phi}{f_a}+\dfrac{n_4\psi}{f_A}\right)\right]\, ,~~~
\end{eqnarray}
with the domain wall numbers $n_i$.
For simplicity, one can take $n_1=n_3=n_4=1$ and $n_2=0$ to obtain the potential
\begin{eqnarray}
V_{\rm mix}(\phi,\psi)=m_a^2(T) f_a^2\left[1-\cos\left(\dfrac{\phi}{f_a}\right)\right]+ m_A^2 f_A^2\left[1-\cos\left(\dfrac{\phi}{f_a}+\dfrac{\psi}{f_A}\right)\right]\, .
\end{eqnarray}
The dynamics and cosmological evolution of the axions in this case are studied in detail in refs.~\cite{Kitajima:2014xla, Daido:2015cba, Ho:2018qur}.
In the mixing, the mass eigenvalues $m_1(T)$ and $m_2(T)$ of the mass eigenstates $a_1$ and $a_2$ will approach to each other under a certain condition, $\rm e.g.$, both the mass and decay constant of the ALP should be much smaller than the QCD axion, and then move away from each other due the cosmic temperature decreasing, which is called the level crossing.
At level crossing, the energy adiabatic transition can occur between these two mass eigenstates when the adiabatic condition is satisfied.
The final QCD axion abundance and isocurvature perturbations can be suppressed by the adiabatic transition between them \cite{Kitajima:2014xla}.
Furthermore, the level crossing can also lead to the domain walls formation, which is a more common phenomenon in the axiverse than previously thought \cite{Daido:2015bva}.
Recently, the heavy QCD axion DM from the avoided level crossing is studied in ref.~\cite{Cyncynates:2023esj}, in which the QCD axion mixes with the sterile axion, leading to an avoided level crossing of their mass eigenstates.
In addition, a novel single/double level crossings between the $Z_{\mathcal N}$ QCD axion and ALP is investigated in ref.~\cite{Li:2023uvt}, which is different from those mentioned above.

\section{Effects of explicit Peccei-Quinn symmetry breaking}
\label{sec_PQ_symmetry}

In this section, we investigate the effects of the explicit PQ symmetry breaking on axion mass mixing in the axiverse with one QCD axion and one ALP, focusing on the dynamics of the axions and their cosmological evolutions, and also the axion relic density. 

\subsection{Explicit Peccei-Quinn symmetry breaking}
 
In this subsection, we first discuss the explicit PQ symmetry breaking effects on QCD axion DM.
Here we consider the following explicit PQ symmetry breaking potential from the quantum gravity effects \cite{Higaki:2016yqk}
\begin{eqnarray}
V_{\cancel{\rm PQ}}(\phi)=m_{\cancel{\rm PQ}}^2f_a^2\left[1-\cos\left(\dfrac{\phi}{f_a}-\theta_H\right)\right]\, ,
\label{V_PQ}
\end{eqnarray}
where $m_{\cancel{\rm PQ}}$ is the corresponding effective mass, and $\theta_H$ is the phase.
Since the strong CP phase $\bar{\theta}$ is strongly constrained by the nEDM measurement $|\bar{\theta}|<10^{-10}$ \cite{Abel:2020pzs}, we have the upper limit on the explicit PQ breaking with
\begin{eqnarray}
R_m\equiv\dfrac{m_{\cancel{\rm PQ}}}{m_{a,0}} < \sqrt{\bigg|\dfrac{10^{-10}}{\sin\left(10^{-10}-\theta_H\right)}\bigg|}\, ,
\end{eqnarray}
which is given by solving $V'_{\rm QCD}(\phi)+V'_{\cancel{\rm PQ}}(\phi)=0$, where the prime represents derivative with respect to the QCD axion field $\phi$.
In this case, the axion starts to oscillate when its mass becomes comparable to the Hubble parameter $H(T)$, $3H(T'_{{\rm osc},a})= m_{\cancel{\rm PQ}}$, with the oscillation temperature
\begin{eqnarray}
T'_{{\rm osc},a}\simeq1.33 \, {\rm GeV} \left(\dfrac{g_*(T'_{{\rm osc},a})}{61.75}\right)^{-1/4} \left(\dfrac{R_m}{10^{-3}}\right)^{1/2} \left(\dfrac{f_a}{10^{12}\, \rm GeV}\right)^{-1/2}\, .
\end{eqnarray}
Since we are more concerned about a case that the QCD axion starts to oscillate before the conventional case, $\rm i.e.$, $T'_{{\rm osc},a} > T_{{\rm osc},a}$, we have the lower limit with
\begin{eqnarray}
R_m > 5.17\times10^{-4} \left(\dfrac{g_*(T'_{{\rm osc},a})}{61.75}\right)^{0.34}\left(\dfrac{f_a}{10^{12}\, \rm GeV}\right)^{0.67}\, .
\label{R_m_limit}
\end{eqnarray}
Then the QCD axion DM abundance in this case is given by
\begin{eqnarray}
\begin{aligned}
\Omega_ah^2&\simeq7.15\times10^{-2} \left(\dfrac{g_{*s}(T_0)}{3.94}\right)\left(\dfrac{g_*(T'_{{\rm osc},a})}{61.75}\right)^{-1/4}\left(\dfrac{R_m}{10^{-3}}\right)^{-1/2}\\
&\times\left(\dfrac{f_a}{10^{12}\, \rm GeV}\right)^{3/2}\left\langle\left(\theta_i-\theta_H\right)^2f'(\theta_i)\right\rangle\, ,
\end{aligned}
\end{eqnarray}
with the anharmonicity factor \cite{Lyth:1991ub, Visinelli:2009zm, Jeong:2022kdr}
\begin{eqnarray}
f'(\theta_i)\simeq\left[\ln\left(\dfrac{e}{1-\left(\theta_i-\theta_H\right)^2/ \pi^2}\right)\right]^{3/2}\, .
\end{eqnarray}
A more complicated case with the multiple minima of the explicit PQ breaking term is discussed in ref.~\cite{Jeong:2022kdr}, in which the trapping effect on QCD axion DM abundance and isocurvature perturbations is significant in the case with a large $|\theta_i-\theta_H|$, depending on around which vacuum the axion first starts to oscillate.
 
\subsection{Effects on axion mass mixing in the axiverse}

In the following, we investigate the effects of the above explicit PQ symmetry breaking on axion mass mixing in the axiverse.
In this case, the axion mixing potential is given as follows with the notation
\begin{eqnarray}
V_{\rm mix}(\phi, \psi)\to V_{\rm mix}(\phi, \psi)+V_{\cancel{\rm PQ}}(\phi)\, .
\end{eqnarray}
Using eq.~(\ref{V_PQ}), we derive the mixing potential in our model
\begin{eqnarray}
\begin{aligned}
V_{\rm mix}(\phi, \psi)&=m_a^2(T)f_a^2\left[1-\cos\left(\dfrac{\phi}{f_a}\right)\right]+m_{\cancel{\rm PQ}}^2f_a^2\left[1-\cos\left(\dfrac{\phi}{f_a}-\theta_H\right)\right]\\
&+ m_A^2 f_A^2\left[1-\cos\left(\dfrac{\phi}{f_a}+\dfrac{\psi}{f_A}-\theta_H\right)\right]\, .
\label{V_mix}
\end{aligned}
\end{eqnarray}
Note that this is a specific potential form, the integer coefficient $\sim N_H(\phi/f_a-\theta_H)$ may also exist in the term. 
For simplicity, here we take $N_H=1$.
For our purpose, it is necessary to eliminate the phase difference between the first two terms on the right side of eq.~(\ref{V_mix}), so we take $\theta_H=0$.
This can be seen as equivalent to requiring a separate solution to the strong CP problem.
Since this choice does not affect the evolution of the two axion fields during the mixing process, it will not influence the ultimate production of axion DM in our scenario.
In addition, see $\rm e.g.$ refs.~\cite{Alonso-Alvarez:2017hsz, DiLuzio:2021gos} for the other similar axion potentials and also the axion DM production.
In ref.~\cite{Alonso-Alvarez:2017hsz}, they discuss the ALP DM with a non-canonical kinetic term, and also coupled to QCD.
In ref.~\cite{DiLuzio:2021gos}, they consider the lighter QCD axion DM production through the novel trapped misalignment mechanism, including the nontrivial temperature dependence of the axion potentials.

\begin{figure}[t]
\centering
\includegraphics[width=0.64\textwidth]{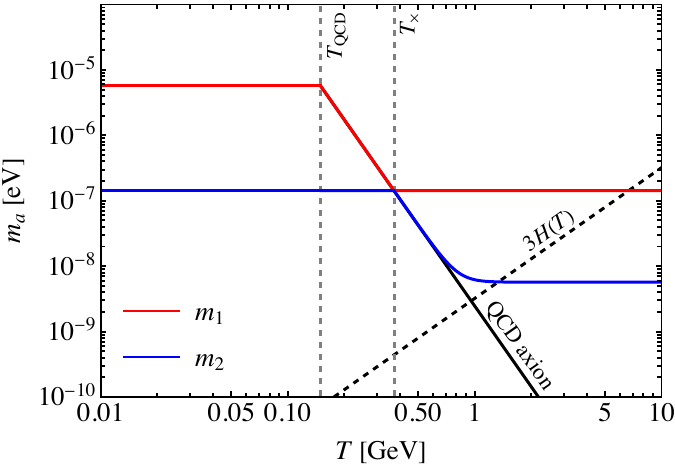}
\caption{The temperature-dependent mass eigenvalues $m_{1,2}(T)$ as functions of the cosmic temperature $T$ with the typical parameter set for level crossing to occur.
The red and blue solid lines represent the mass eigenvalues $m_1(T)$ and $m_2(T)$, respectively.
The black solid line represents the QCD axion mass $m_a(T)$.
The black dashed line represents the Hubble parameter $H(T)$.
The two vertical gray lines represent the temperatures $T_{\rm QCD}$ and $T_\times$, respectively.
Here we set $f_a=10^{12}\, \rm GeV$, $R_m=0.001$, $r_m=0.025$, and $r_f=0.01$.}
\label{fig_m12}
\end{figure}

Substituting eq.~(\ref{V_mix}) into the equations of motion (EOM)
\begin{eqnarray}
\ddot\phi+3H\dot\phi+\dfrac{\partial V_{\rm mix}(\phi, \psi)}{\partial \phi}=0 \, , \quad  \ddot\psi+3H\dot\psi+\dfrac{\partial V_{\rm mix}(\phi, \psi)}{\partial \psi}=0 \, ,
\end{eqnarray}
we derive the EOM of $\phi$ and $\psi$ 
\begin{eqnarray}
\ddot\phi+3H\dot\phi+\left(m_a^2(T)+m_{\cancel{\rm PQ}}^2\right)f_a\sin\left(\dfrac{\phi}{f_a}\right)+m_A^2 \dfrac{f_A^2}{f_a} \sin\left(\dfrac{\phi}{f_a}+\dfrac{\psi}{f_A}\right)=0\, , 
\end{eqnarray}
\begin{eqnarray}
\ddot\psi+3H\dot\psi+m_A^2 f_A\sin\left(\dfrac{\phi}{f_a}+\dfrac{\psi}{f_A}\right)=0 \, , 
\end{eqnarray}
where the dots represent derivatives with respect to the physical time $t$.
Considering the oscillation amplitudes of $\phi$ and $\psi$ are much smaller than the corresponding decay constants, we have the mass mixing matrix
\begin{eqnarray}
\mathbf{M}^2=
\left(
\begin{array}{cc}
m_a^2(T)+m_{\cancel{\rm PQ}}^2+m_A^2\dfrac{f_A^2}{f_a^2}  & \quad m_A^2\dfrac{f_A}{f_a} \\
m_A^2\dfrac{f_A}{f_a} & \quad   m_A^2
\end{array}
\right)\, .
\end{eqnarray}
Diagonalizing the mass mixing matrix, we derive the mass eigenstates $a_1$ and $a_2$
\begin{eqnarray}
\left(
\begin{array}{c}
a_1 \\
a_2 
\end{array}
\right)
=
\left(
\begin{array}{cc}
\cos \alpha & \quad \sin \alpha \\
-\sin \alpha  & \quad   \cos \alpha
\end{array}
\right)
\left(
\begin{array}{c}
\phi \\
\psi 
\end{array}
\right)\, ,
\end{eqnarray} 
with the mass mixing angle $\alpha$
\begin{eqnarray}
\cos^2\alpha(T)=\dfrac{1}{2}\left(1+\dfrac{m_A^2-m_a^2(T)-m_{\cancel{\rm PQ}}^2-m_A^2\dfrac{f_A^2}{f_a^2}}{m_1^2(T)-m_2^2(T)}\right) \, ,
\end{eqnarray} 
where $m_1(T)$ and $m_2(T)$ ($m_1(T)>m_2(T)$) are the corresponding mass eigenvalues
\begin{eqnarray}
\begin{aligned}
m_{1,2}^2(T)&=\dfrac{1}{2}\left[m_a^2(T)+m_{\cancel{\rm PQ}}^2+m_A^2+m_A^2 \dfrac{f_A^2}{f_a^2}\right]\\
&\pm\dfrac{1}{2f_a^2}\bigg[-4 m_A^2 f_a^4 \left(m_a^2(T)+m_{\cancel{\rm PQ}}^2\right)\\
&+\left(m_A^2 f_A^2+f_a^2 \left(m_a^2(T)+m_{\cancel{\rm PQ}}^2+m_A^2\right)\right)^2\bigg]^{1/2}\, .
\end{aligned}
\end{eqnarray}  
Then we discuss the level crossing in the mixing.
The timing of the level crossing is that the difference of $m_1^2(T)-m_2^2(T)$ gets a minimum value at the level crossing temperature $T_\times$, $\rm i.e.$, ${\rm d}\left(m_1^2(T)-m_2^2(T)\right)/{\rm d}T=0$, which in our scenario is given by
\begin{eqnarray}
T_\times= T_{\rm QCD}\left(m_{a,0}\right)^{1/4.08}\left(m_A^2- m_{\cancel{\rm PQ}}^2-\dfrac{m_A^2 f_A^2}{f_a^2}\right)^{-1/8.16}\, .   
\end{eqnarray} 
In this case, the QCD axion mass at $T_\times$ reads
\begin{eqnarray}
m_{a,\times}\equiv m_a(T_\times)\simeq \sqrt{m_A^2- m_{\cancel{\rm PQ}}^2-\dfrac{m_A^2 f_A^2}{f_a^2}}\, .   
\end{eqnarray} 
Using $0<m_a(T_\times)/m_{a,0}<1$, we derive the condition for level crossing
\begin{eqnarray}
0 < r_m^2-R_m^2- r_m^2 r_f^2 < 1\, ,
\label{clc_1}
\end{eqnarray} 
where we have defined $r_m\equiv m_A/m_{a,0}$ and $r_f\equiv f_A/f_a$.
We show in figure~\ref{fig_m12} the temperature-dependent mass eigenvalues $m_{1,2}(T)$ as functions of the cosmic temperature $T$ with the typical parameter set for level crossing to occur.
Since we are more interested in the parameters $r_m\ll1$ and $r_f\ll1$, here we set $f_a=10^{12}\, \rm GeV$, $R_m=0.001$, $r_m=0.025$, and $r_f=0.01$.
The red and blue lines correspond to the mass eigenvalues $m_1(T)$ and $m_2(T)$, respectively.
The black solid and dashed lines correspond to the temperature-dependent QCD axion mass $m_a(T)$ [eq.~(\ref{maT})] and the Hubble parameter $H(T)$, respectively.
We also show the level crossing temperature $T_\times$ and $T_{\rm QCD}$ in the plot.

Let us discuss the temperature-dependent behavior of the axions when the mass mixing occurs in figure~\ref{fig_m12}. 
At high temperatures, the light mass eigenstate $a_2$ comprises the QCD axion, while the heavy mass eigenstate $a_1$ comprises the ALP.
With the cosmic temperature decreasing, the level crossing will take place, the masse eigenvalues $m_1(T)$ and $m_2(T)$ will approach to each other at the temperature $T_\times$ and then move away from each other.
After that, the $a_2$ comprises the ALP, and the $a_1$ comprises the QCD axion.
The level crossing can lead to the adiabatic energy density transition between the two mass eigenstates, or between the QCD axion and ALP, which will be discussed next.
Since $r_m$ determines the ALP mass $m_A$, it will mainly affect the temperature $T_\times$.
For different values of $r_f$, the main difference is their behaviors at $T_\times$, which maybe the effect of avoided level crossing for larger $r_f$ as discussed in ref.~\cite{Cyncynates:2023esj}, but we will not discuss it further in this text.
In order for the level crossing to occur, we note that there is an additional relation that should be satisfied, the value of $m_2(T)$ at high temperatures should not be larger than the ALP mass, $\rm i.e.$, $R_m < r_m$.
There is no level crossing in the case $R_m > r_m$, which can be seen from figure~\ref{fig_m12}.
However, we find that this relation is already given in eq.~(\ref{clc_1}) with $r_m^2-R_m^2>r_m^2 r_f^2 > 0$.
 
\begin{figure}[t]
\centering
\includegraphics[width=0.64\textwidth]{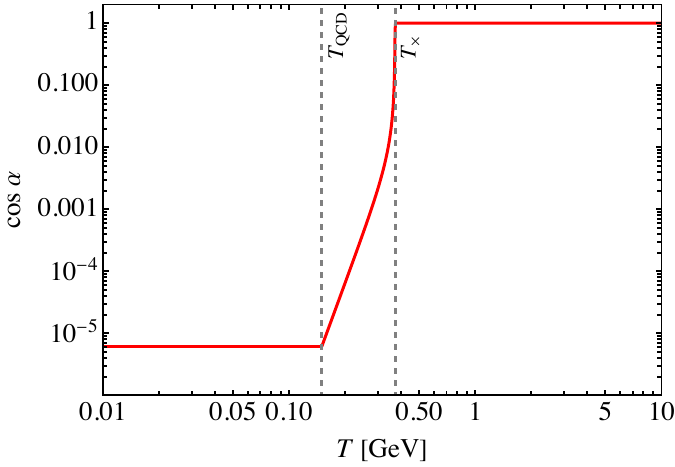}
\caption{The mass mixing angle $\cos\alpha$ as a function of $T$ with the parameter set in figure~\ref{fig_m12}.}
\label{fig_angle}
\end{figure} 

In addition, we also show the mass mixing angle $\cos\alpha$ as a function of the cosmic temperature in figure~\ref{fig_angle}, in which the parameter set corresponds to figure~\ref{fig_m12}.
At high temperatures, we have the constant value of the mixing angle with $\cos\alpha=1$ ($\alpha=0$) until the level crossing occurs at the temperature $T_\times$.
Then the $\cos\alpha$ drops rapidly and stabilizes at a small value, which lasts for a very short time, $\Delta t_\times$.
Note that the behaviors of the mass mixing angle are not always same as shown in figure~\ref{fig_angle}, which is exactly the case for level crossing to occur.
In fact, the alteration of the mixing angle is not significant or even unchanged if there is no level crossing in the mixing, which is not the interest of this text and we do not show it.

\begin{figure}[t]
\centering
\includegraphics[width=0.64\textwidth]{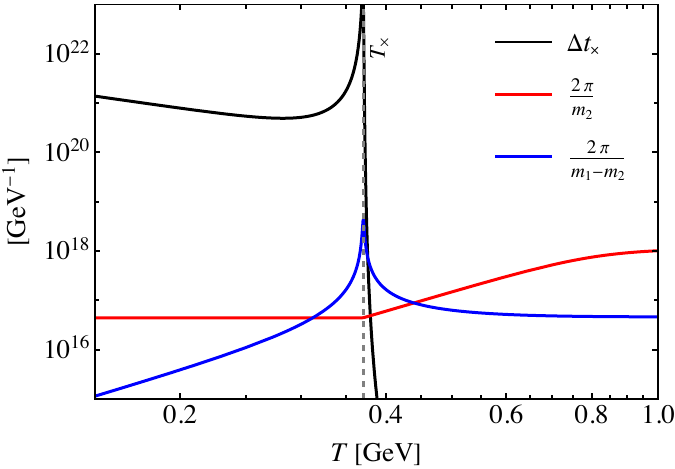}
\caption{The three terms in eq.~(\ref{adiabatic}) as functions of $T$ with the parameter set in figure~\ref{fig_m12}.
The black, red, and blue lines represent the values of $\Delta t_\times$, $2\pi/m_2(T)$, and $2\pi/\left(m_1(T)-m_2(T)\right)$, respectively.
The vertical gray line represents the temperature $T_\times$.}
\label{fig_deltatx}
\end{figure}

Then we briefly discuss the adiabatic energy density transition between the two mass eigenstates in our scenario, it is necessary to determine the axion energy density.
In the mass mixing, the adiabatic transition will take place if both the comoving axion numbers of the eigenstates $a_1$ and $a_2$ are separately conserved at the level crossing \cite{Ho:2018qur, devaud:hal-00197565}, which lasts for a parametric duration
\begin{eqnarray}
\Delta T_\times=\bigg|\dfrac{1}{\cos\alpha(T)}\dfrac{{\rm d}\cos\alpha(T)}{{\rm d} T}\bigg|^{-1}_{T=T_\times}\, ,
\end{eqnarray}  
corresponding to the time $\Delta t_\times$.
The condition for adiabatic transition reads
\begin{eqnarray}
\Delta t_\times \gg \max\left[\dfrac{2\pi}{m_2(T)}\bigg|_{T=T_\times}, \, \dfrac{2\pi}{m_1(T)-m_2(T)}\bigg|_{T=T_\times}\right]\, ,
\label{adiabatic_transition}
\label{adiabatic}
\end{eqnarray}
which has a complicated analytical form, and here we just check it numerically. 
We show the values of the three terms in eq.~(\ref{adiabatic}) in figure~\ref{fig_deltatx}.
The parameter set also corresponds to figure~\ref{fig_m12}.
The black, red, and blue lines correspond to the values of $\Delta t_\times$, $2\pi/m_2(T)$, and $2\pi/\left(m_1(T)-m_2(T)\right)$, respectively.
From this plot, we find that eq.~(\ref{adiabatic_transition}) can be satisfied at the level crossing $T_\times$.

\subsection{Axion relic density}

In this subsection, we estimate the relic density of the QCD axion and ALP DM in our scenario through the misalignment mechanism.
We still consider the pre-inflationary scenario in which the PQ symmetry is spontaneously broken during inflation.

We first discuss the QCD axion DM.
At high temperatures, the ALP field is frozen at an arbitrary initial misalignment angle $\theta_{i,A}$ and starts to oscillate at $T_{{\rm osc},A}$, we have the initial energy density in the ALP field
\begin{eqnarray}
\rho_{A,{\rm osc}}=\frac{1}{2}m_A^2 f_A^2 \theta_{i,A}^2\, .
\end{eqnarray}
Then at $T_\times<T<T_{{\rm osc},A}$, the ALP energy density is adiabatic invariant with the comoving number $N_A \equiv \rho_A a^3 /m_A$, where $a$ is the scale factor.
The ALP energy density at the level crossing temperature $T_\times$ is given by
\begin{eqnarray}
\rho_{A,\times}=\frac{1}{2}m_A^2 f_A^2 \theta_{i,A}^2 \left(\frac{a_{{\rm osc},A}}{a_\times}\right)^3 \, ,
\end{eqnarray}
where $a_{{\rm osc},A}$ and $a_\times$ correspond to the scale factors at $T_{{\rm osc},A}$ and $T_\times$, respectively.
At the level crossing, the heavy mass eigenstate $a_1$ will comprise the QCD axion and this energy density $\rho_{A,\times}$ is transferred to the QCD axion $\rho_{a,\times}$.
At $T<T_\times$, the adiabatic approximation is valid again with $N_a \equiv \rho_a a^3 /m_a$.
Using 
\begin{eqnarray}
\dfrac{\rho_{a,\times} a_\times^3}{m_{a,\times}}=\dfrac{\rho_{a,0} a_0^3}{m_{a,0}}\, ,
\end{eqnarray}
we have the current QCD axion energy density
\begin{eqnarray}
\rho_{a,0}=\frac{1}{2}m_{a,0}m_A f_A^2 \theta_{i,A}^2 \left(\frac{a_{{\rm osc},A}}{a_0}\right)^3 \, ,
\end{eqnarray}
where $a_0$ is the scale factor today.
Compared with the no level crossing case
\begin{eqnarray}
\rho'_{a,0}=\frac{1}{2}m_{a,0} m_{a,{\rm osc}} f_a^2 \theta_{i,a}^2 \left(\frac{a_{{\rm osc},a}}{a_0}\right)^3 \, ,
\end{eqnarray}
we find that the QCD axion relic density can be suppressed by a factor 
\begin{eqnarray}
\sqrt{m_{a,{\rm osc}}/m_A}r_f^2 \ll 1\, ,
\end{eqnarray}
where we ignore the difference between the initial misalignment angles.
Note that $m_{a,{\rm osc}}\equiv m_a(T_{{\rm osc},a})$, corresponding to eq.~(\ref{Tosca}).

Next we discuss the ALP DM in a similar way.
At high temperatures, the QCD axion starts to oscillate at $T'_{{\rm osc},a}$ with the initial misalignment angle $\theta_{i,a}$, and the initial QCD axion energy density is given by 
\begin{eqnarray}
\rho_{a,{\rm osc'}}=\frac{1}{2}m_{\cancel{\rm PQ}}^2 f_a^2 \theta_{i,a}^2\, .
\end{eqnarray}
At $T_\times<T<T'_{{\rm osc},a}$, the QCD axion energy density is adiabatic invariant, which at $T_\times$ can be described by
\begin{eqnarray}
\rho_{a,\times}=\frac{1}{2} m_A m_{\cancel{\rm PQ}} f_a^2 \theta_{i,a}^2 \left(\frac{a_{{\rm osc'},a}}{a_\times}\right)^3 \, ,
\end{eqnarray}
where $a_{{\rm osc'},a}$ is the scale factor at $T'_{{\rm osc},a}$.
Then at the level crossing $T=T_\times$, the light mass eigenstate $a_2$ will comprise the ALP and this energy density is transferred to the ALP $\rho_{A,\times}$.
At $T<T_\times$, the adiabatic approximation is valid again, then we have the current ALP energy density 
\begin{eqnarray}
\rho_{A,0}=\frac{1}{2}m_A m_{\cancel{\rm PQ}} f_a^2 \theta_{i,a}^2 \left(\frac{a_{{\rm osc'},a}}{a_0}\right)^3 \, .
\end{eqnarray}
Compared with the no level crossing case
\begin{eqnarray}
\rho'_{A,0}=\frac{1}{2}m_A^2 f_A^2 \theta_{i,A}^2 \left(\frac{a_{{\rm osc},A}}{a_0}\right)^3 \, ,
\end{eqnarray}
we find that the ALP relic density can be enhanced by a factor 
\begin{eqnarray}
\sqrt{m_A/m_{\cancel{\rm PQ}}}/r_f^2=\sqrt{r_m/R_m}/r_f^2 \gg 1\, .
\end{eqnarray}

\section{Conclusion}
\label{sec_Conclusion}

In this paper, we have investigated the axion mass mixing between the QCD axions and ALPs in the axiverse with the explicit PQ symmetry breaking in the early Universe. 
We first review the QCD axion and ALP DM through the misalignment mechanism, and introduce the mass mixing between them.
Then we investigate this mass mixing with an explicit PQ symmetry breaking term.
Exactly, we consider the mass mixing between one QCD axion and one ALP in this work.
We study the dynamics of the axions and their cosmological evolutions in detail when the level crossing occurs in this scenario.
We show the evolution of the mass eigenvalues and the mass mixing angle with the typical parameter set for level crossing to occur.
Then we check the condition for energy adiabatic transition with the corresponding parameter set, which is necessary to determine the axion energy density.
Finally, we estimate the QCD axion and ALP relic density through the misalignment mechanism.
We find that, the QCD axion relic density can be suppressed by a factor $\sim \sqrt{m_{a,{\rm osc}}/m_A}r_f^2$, while the ALP relic density can be enhanced by a factor $\sim \sqrt{r_m/R_m}/r_f^2$.
The level crossing in our scenario may also have some interesting cosmological implications, such as the axion domain walls formation around the level crossing, the nano-Hertz gravitational waves emitted by the domain walls annihilation, and the primordial black holes formation from the domain walls collapse.
 
\section*{Acknowledgments}

The author would like to thank Wei Chao, David Cyncynates, Shota Nakagawa, Ying-Quan Peng, Wen Yin, and Yu-Feng Zhou for helpful discussions and valuable comments.
This work was partly supported by the Key Laboratory of Theoretical Physics in Institute of Theoretical Physics, CAS.


\bibliographystyle{JHEP}
\bibliography{references}

\end{document}